\documentclass[aps,preprintnumbers,nofootinbibt,referee, keywords]{revtex4}
\usepackage{graphicx}
\usepackage{amsmath}
\usepackage{bm}
\usepackage{color}

\def\be{\begin{equation}}
\def\ee{\end{equation}}
\def\bea{\begin{eqnarray}}
\def\eea{\end{eqnarray}}

\begin{document}

\title{New further integrability cases for the Riccati equation}
\author{M. K. Mak}
\email{mkmak@vtc.edu.hk}
\affiliation{Department of Physics and Center for Theoretical and Computational Physics,
The University of Hong Kong, Pok Fu Lam Road, Hong Kong, P. R. China}
\author{T. Harko}
\email{t.harko@ucl.ac.uk}
\affiliation{Department of Mathematics, University College London, Gower Street, London WC1E 6BT, United Kingdom}

\begin{abstract}
New further integrability conditions of the Riccati equation $%
dy/dx=a(x)+b(x)y+c(x)y^{2}$ are presented. The first case corresponds to
fixed functional forms of the coefficients $a(x)$ and $c(x)$ of the Riccati
equation, and of the function $F(x)=a(x)+\left[f(x)-b^{2}(x)\right]/4c(x)$,
where $f(x)$ is an arbitrary function. The second integrability case is
obtained for the "reduced" Riccati equation with $b(x)\equiv 0$. If the
coefficients $a(x)$ and $c(x)$ satisfy the condition $\pm d\sqrt{f(x)/c(x)}%
/dx=a(x)+f(x)$, where $f(x)$ is an arbitrary function, then the general
solution of the "reduced" Riccati equation can be obtained by quadratures.
The applications of the integrability condition of the "reduced" Riccati
equation for the integration of the Schr\"odinger and Navier-Stokes
equations are briefly discussed.

\end{abstract}

\maketitle

%\pacs{67.85.Jk, 04.40.Dg, 95.30.Cq, 95.30.Sf}

\section{ Introduction}

Recently two integrability conditions of the Riccati equation
\begin{equation}
\frac{dy}{dx}=a(x)+b(x)y+c(x)y^{2},  \label{1}
\end{equation}%
where $a$, $b$, $c$ are arbitrary real functions of $x$, with $a,b,c\in
C^{\infty }(I)$, defined on a real interval $I\subseteq \Re $ \cite{1r}, %
\cite{2r}, have been obtained in \cite{AMC}. By introducing a solution
generating function $f(x)\in C^{\infty }(I)$, one can attach to Eq.~(\ref{1}%
) the auxiliary Riccati equation
\begin{equation}
\frac{dy}{dx}=\frac{d}{dx}\left[ \frac{-b(x)\pm \sqrt{f(x)}}{2c(x)}\right] +%
\frac{b^{2}(x)-f(x)}{4c(x)}+b(x)y+c(x)y^{2}.  \label{2}
\end{equation}%
Then, if the coefficients $a$, $b$, $c$ of the Riccati Eq.~(\ref{1}) satisfy
the condition
\begin{equation}  \label{4_1}
a(x)=\frac{d}{dx}\left[ \frac{-b(x)\pm \sqrt{f(x)}}{2c(x)}\right] +\frac{%
b^{2}(x)-f(x)}{4c(x)}.
\end{equation}
the general solution of Eq.~(\ref{2}) is obtained as
\begin{equation}
y_{\pm }(x)=e^{\pm \int ^x\sqrt{f\left(x'\right)}dx'}\left[ -\int ^x c\left(x'\right)e^{\pm \int ^{x'}\sqrt{f\left(x''\right)%
}dx''}dx'+C_{\pm }\right] ^{-1}+\left[ \frac{-b(x)\pm \sqrt{f(x)}}{2c(x)}\right]
,  \label{6_1}
\end{equation}%
where $C_{\pm }$ are arbitrary integration constants \cite{AMC}.

In \cite{AMC} the general solution of the Riccati Eq.~(\ref{2}) was obtained
and discussed in detail in two cases. First, the functional forms of the
functions $b(x)$, $c(x)$ and $f(x)$, was fixed, and the general solution of
the Riccati equation was obtained with $a(x)$ given by Eq.~(\ref{4_1}). In
the second case, by fixing the functional forms of the functions $a(x)$, $%
b(x)$ and $f(x)$, the general solution of the Riccati Eq.~(\ref{2}) was
obtained with $c(x)$ given by Eq.~(\ref{4_1}).

It is the purpose of the present paper to consider some further
integrability conditions of the Riccati Eq.~(\ref{1}), by using the approach
introduced in \cite{AMC}. An integrability condition can be obtained by
fixing the functional forms of the functions $a(x)$, $c(x)$, and of the
function $F(x)=a(x)+\left[ f(x)-b^{2}(x)\right] /4c(x)$, which allows us to
obtain the general solution of the Riccati Eq.~(\ref{2}), with $b(x)$
determined from Eq.~(\ref{4_1}).

An integrability case for the "reduced" Riccati equation with $b(x)\equiv 0$
is also obtained. Some physical applications of the solution are considered
for the case of the Schr\"odinger equation, and of the Navier-Stokes
equation, respectively.

The present paper is organized as follows. The general solution of the
Riccati equation with given $a$, $c$ and $F$ is derived in Section~\ref%
{sect1}. An integrability case for the "reduced" Riccati equation is
presented in Section~\ref{sect2}. Some physical applications of the solution
generating method for the "reduced" Riccati equation are presented in
Section~\ref{sect3}. We conclude our results in Section~\ref{sect4}.

\section{Generating solutions of the Riccati equation by fixing $a(x)$, $%
c(x) $ and $F(x)$}

\label{sect1}

Eq.~(\ref{4_1}) can be rewritten as
\begin{equation}
F(x)=\frac{d}{dx}\left[ \frac{-b(x)\pm \sqrt{f(x)}}{2c(x)}\right] ,
\label{b11}
\end{equation}
where we have introduced a new generating function $F\left( x\right) $
defined as
\begin{equation}
F(x)=a(x)+\frac{f(x)-b^{2}(x)}{4c(x)}.  \label{F}
\end{equation}

From Eq.~(\ref{F}) we obtain $f(x)$ as
\begin{equation}
f(x)=4c(x)\left[ F(x)-a(x)\right] +b^{2}(x).  \label{f}
\end{equation}

Integrating Eq.~(\ref{b11}) yields the following equation:
\begin{equation}
2c(x)\left( \int ^xF\left(x'\right)dx'+F_{0}\right) +b(x)=\pm \sqrt{f(x)},  \label{Ff}
\end{equation}%
where $F_{0}$ is an arbitrary integration constant. By inserting Eq.~(\ref{f}%
) into Eq.~(\ref{Ff}), we find
\begin{equation}
2c(x)\left[ \int ^x F\left(x'\right)dx'+F_{0}\right] +b(x)=\pm \sqrt{4c(x)\left[ F(x)-a(x)%
\right] +b^{2}(x)}.  \label{FfN}
\end{equation}

By squaring both sides of Eq.~(\ref{FfN}), we obtain the following equation:
\begin{equation}
\left[ 2c(x)\left( \int ^x F\left(x'\right)dx'+F_{0}\right) +b(x)\right] ^{2}=4c(x)\left[
F(x)-a(x)\right] +b^{2}(x).  \label{Squ}
\end{equation}

Now rearranging the terms of Eq.~(\ref{Squ}), it is easy to show that
\begin{equation}
b(x)=\frac{F(x)-a(x)}{\int ^x F\left(x'\right)dx'+F_{0}}-c(x)\left[ \int ^x F\left(x'\right)dx'+F_{0}\right]
.  \label{bbb}
\end{equation}

By substituting Eq.~(\ref{bbb}) into Eq.~(\ref{Ff}), Eq.~(\ref{Ff}) can be
expressed as
\begin{equation}
\pm \sqrt{f(x)}=c(x)\left[ \int ^x F\left(x'\right)dx'+F_{0}\right] +\frac{F(x)-a(x)}{\int ^x
F\left(x'\right)dx'+F_{0}}.  \label{FFF}
\end{equation}

By substituting Eq.~(\ref{bbb}) into the Riccati Eq.~(\ref{1}), the Riccati
Eq.~(\ref{1}) takes the form
\begin{equation}
\frac{dy}{dx}=a(x)+\left\{ \frac{F(x)-a(x)}{\int ^xF\left(x'\right)dx'+F_{0}}-c(x)\left[
\int ^x F\left(x'\right)dx'+F_{0}\right] \right\} y+c(x)y^{2}.  \label{bR}
\end{equation}

Therefore we obtain the following:

\textbf{Theorem}. Let $F(x)\in C^{\infty }(I)$ be an arbitrary function
defined on a real interval $I\subseteq \Re $. If the coefficient $b(x)$ of
the Riccati Eq.~(\ref{1}) satisfies the condition (\ref{bbb}), then the
general solution of the Riccati Eq.~(\ref{bR}) is given by
\begin{equation}
y(x)=\frac{e^{\int ^x \left\{ c\left(x'\right)\left[ \int ^{x'}F\left(x''\right)dx''+F_{0}\right] +\frac{%
F\left(x'\right)-a\left(x'\right)}{\int ^{x'} F\left(x''\right)dx''+F_{0}}\right\} dx'}}{C_{0}-\int ^xc(x')e^{\int ^{x'} \left\{
c\left(x''\right)\left[ \int ^{x''} F(\xi )d\xi+F_{0}\right] +\frac{F\left(x''\right)-a\left(x''\right)}{\int ^{x''}F\left(\xi \right)d\xi +F_{0}}%
\right\} dx''}dx'}+\int ^xF\left(x'\right)dx'+F_{0},  \label{yR}
\end{equation}
where $C_{0}$ is an arbitrary integration constant.

\subsection{Example 1}

The coefficients of the Riccati equation
\begin{equation}  \label{ex1}
\frac{dy}{dx}=k_{1}x^{m}-k_{2}x^{n}\left( F_{0}+\frac{k_{1}x^{m+1}}{m+1}%
\right) y+k_{2}x^{n}y^{2},
\end{equation}
where $k_{1}$, $k_{2}$, $m$ and $n$ are arbitrary constants, satisfy the
condition (\ref{bbb}). A particular solution of Eq.~(\ref{ex1}) is
\begin{equation}
y_{p}=F_{0}+\frac{k_{1}x^{1+m}}{1+m}.
\end{equation}%
The general solution of Eq.~(\ref{ex1}) is given by
\begin{equation}
y(x)=F_{0}+\frac{k_{1}x^{m+1}}{m+1}+\frac{\exp \left\{ {k_{2}x^{n+1}\left[
\frac{k_{1}x^{m+1}}{(m+1)(m+n+2)}+\frac{F_{0}}{n+1}\right] }\right\} }{%
C_{0}-k_{2}\int ^x \xi ^{n}\exp \left\{ {k_{2}\xi ^{n+1}\left[ \frac{k_{1}\xi ^{m+1}}{%
(m+1)(m+n+2)}+\frac{F_{0}}{n+1}\right] }\right\} d\xi }.
\end{equation}

\subsection{Example 2}

The coefficients of the Riccati equation
\begin{equation}  \label{ex2}
\frac{dy}{dx}=k_{1}x+\left[ \frac{\left( n-k_{1}\right) x}{nx^{2}/2+F_{0}}-1%
\right] y+\frac{1}{nx^{2}/2+F_{0}}y^{2},
\end{equation}
satisfy the condition (\ref{bbb}). It has the particular solution
\begin{equation}
y_{p}=F_{0}+\frac{nx^{2}}{2}.
\end{equation}%
The general solution of Eq.~(\ref{ex2}) is given by
\begin{equation}
y(x)=\frac{e^{x}\left( nx^{2}+2F_{0}\right) ^{1-k_{1}/n}}{C_{0}-2\int ^x
e^{\xi }\left( n\xi ^{2}+2F_{0}\right) ^{-k_{1}/n}d\xi }+\frac{nx^{2}}{2}+F_{0}.
\end{equation}

\subsection{Example 3}

The coefficients of the Riccati equation
\begin{equation}
\frac{dy}{dx}=kx^{m}+\left[ \frac{2\left( px^{s}-kx^{m}\right) }{%
px^{s+1}/(s+1)+F_{0}}-\frac{n}{x}\right] y+\left[ \frac{n}{x}-\frac{%
px^{s}-kx^{m}}{px^{s+1}/(s+1)+F_{0}}\right] \left[ \frac{px^{s+1}}{s+1}+F_{0}%
\right] ^{-1}y^{2},  \label{ex3}
\end{equation}
where $k$, $m$, $n$, $s$ and $p$ are arbitrary constants, satisfy the
condition (\ref{bbb}). It has the particular solution
\begin{equation}
y_{p}=F_{0}+\frac{px^{s+1}}{s+1}.
\end{equation}

The general solution of Eq.~(\ref{ex3}) is given by
\begin{equation}
y\left( x\right) =-G\left( x\right) +\frac{x^{n}}{C_{0}-G\left( x\right) },
\end{equation}
where the function $G\left( x\right) $ is defined as
\begin{equation}
G\left( x\right) =\frac{x^{n}}{F_{0}^{2}(s+1)}\left[ \frac{F_{0}(s+1)\left(
kx^{m+1}+F_{0}+F_{0}s\right) }{px^{s+1}+F_{0}+F_{0}s}-\frac{%
k(m+n-s)x^{m+1}\;_{2}F_{1}\left( 1,\frac{m+n+1}{s+1};\frac{m+n+s+2}{s+1};-%
\frac{px^{s+1}}{sF_{0}+F_{0}}\right) }{m+n+1}\right],
\end{equation}
where $_{2}F_{1}\left( 1,\frac{m+n+1}{s+1};\frac{m+n+s+2}{s+1};-\frac{%
px^{s+1}}{sF_{0}+F_{0}}\right) $ is the hypergeometric function $%
_{2}F_{1}\left( a,b;c;z\right) $ \cite{Abram}.

\section{The "reduced" Riccati equation: the case $b(x)\equiv 0$}

\label{sect2}

If the function $b(x)\equiv 0$, the Riccati Eq.~(\ref{1}) takes the
"reduced" form
\begin{equation}
\frac{dy}{dx}=a(x)+c(x)y^{2},  \label{b1}
\end{equation}

We assume that a particular solution $y_p(x)$ of Eq.~(\ref{b1}) satisfies
the condition
\begin{equation}  \label{b2}
\frac{dy_p}{dx}=a(x)+f(x),
\end{equation}
where $f(x)\in C^{\infty }(I)$ is an arbitrary function. Substitution of
Eq.~(\ref{b2}) into Eq.~(\ref{b1}) fixes the particular solution $y_p(x)$ as
\begin{equation}
y_p(x)=\pm\sqrt{\frac{f(x)}{c(x)}}.
\end{equation}

Therefore we obtain an integrability condition of the reduced Riccati
equation, alternative to Eq.~(\ref{4_1}), and which can be formulated as
\begin{equation}  \label{b3}
\pm\frac{d}{dx}\sqrt{\frac{f(x)}{c(x)}}=a(x)+f(x).
\end{equation}

Therefore we have obtained an integrability condition for the "reduced"
Riccati equation expressed as the following

\textbf{Theorem}. Let $f(x)\in C^{\infty }(I)$ be an arbitrary function
defined on a real interval $I\subseteq \Re $. Then the general solution of
the ''reduced'' Riccati equation
\begin{equation}
\frac{dy}{dx}=\pm \frac{d}{dx}\sqrt{\frac{f(x)}{c(x)}}-f(x)+c(x)y^{2},
\label{th1}
\end{equation}%
is given by
\begin{equation}
y_{\pm }(x)=\pm \sqrt{\frac{f(x)}{c(x)}}+\frac{e^{\pm 2\int ^x\sqrt{f\left(x'\right)c\left(x'\right)}%
dx'}}{C_{\pm }-\int ^x c\left(x'\right)e^{\pm 2\int ^{x'} \sqrt{f\left(x''\right)c\left(x''\right)}dx''}dx'},  \label{th2}
\end{equation}%
where $C_{\pm }$ are arbitrary integration constants. If $c(x)\equiv 1$,
the solution of Eq.~(\ref{th2}) can be written as
\begin{equation}
y_{\pm }(x)=\pm \sqrt{f(x)}-\frac{d}{dx}\ln \left[ C_{\pm }-\int ^x{e^{\pm
2\int ^{x'}\sqrt{f\left(x''\right)}dx''}dx'}\right] .
\end{equation}

\section{Applications in physics}

\label{sect3}

In the present Section we consider some physical applications of the
obtained integrability conditions of the Riccati equation. In particular, we
will consider the application of the obtained integrability conditions to
the case of the Schr\"odinger - Riccati system, and of the Navier-Stokes
equations.

\subsection{The Schr\"odinger-Riccati system}

The one-dimensional Schr\"odinger equation for a potential $V (x)$,
\begin{equation}
\psi ^{\prime\prime}(x)+\left[E-V(x)\right]\psi (x)=0,
\end{equation}
where $E=$ constant is the energy, the Planck constant $\hbar =1$, and the
mass $m$ of the particle is normalized to $m=1/2$, by means of the
transformation
\begin{equation}
u(x)=-\frac{\psi ^{\prime}(x)}{\psi (x)},
\end{equation}
can be transformed to a "reduced" Riccati equation of the form \cite{Schr}
\begin{equation}  \label{s1}
u^{\prime}(x)=E-V(x)+u^2(x).
\end{equation}

Therefore if the solution of the ''reduced'' Riccati equation (\ref{s1}) is
known, the wave function is given by
\begin{equation}
\psi (x)=\psi _{0}\exp \left[ -\int ^x {u\left( x^{\prime }\right) dx^{\prime }}%
\right] ,  \label{phi}
\end{equation}%
where $\psi _{0}$ is an arbitrary constant of integration.

For the Schr\"odinger - Riccati equation (\ref{s1}) $c(x)\equiv 1$ and $%
a(x)=E-V(x)$. Therefore for any given arbitrary function $f(x)$ the
Schr\"odinger - Riccati equation has a closed form solution, with the
potential fixed by
\begin{equation}
V(x)=E\mp\frac{d}{dx}\sqrt{f(x)}+f(x).
\end{equation}

As a first application of the integrability case given by Eq.~(\ref{b3}) we
consider the choice $f(x)=f_{0}x^{n}$ for the function $f(x)$, where $f_{0}$
and $n$ are arbitrary constants. Then the potential for this problem is
given by
\begin{equation}
V(x)=E+f_{0}x^{n}\mp \frac{n}{2}\sqrt{f_{0}}x^{n/2-1}.
\end{equation}%
and the general solution of the Schr\"{o}dinger - Riccati Eq.~(\ref{s1}) is
\begin{equation}
u_{\pm }(x)=\pm \sqrt{f_{0}}x^{n/2}+\frac{(n+2)e^{\pm \frac{4\sqrt{f_{0}}}{%
n+2}x^{n/2+1}}}{(n+2)C_{\pm }+2xE_{\frac{n}{n+2}}\left( \mp \frac{4\sqrt{%
f_{0}}}{n+2}x^{n/2+1}\right) },
\end{equation}%
where $E_{n}(z)=\int_{1}^{\infty }{e^{-zt}dt/t^{n}}$ is the exponential
integral function \cite{Abram}. The wave-function corresponding to this
potential is given by
\begin{equation}
\psi _{\pm }\left( x\right) =\psi _{\pm 0}\exp \left[ \mp 2\sqrt{f_{0}}%
x^{n/2+1}/(n+2)\right]\left[ C_{\pm }+2xE_{\frac{n}{n+2}}\left( \mp \frac{4%
\sqrt{f_{0}}}{n+2}x^{n/2+1}\right) /(n+2)\right].
\end{equation}
where $\psi _{\pm 0}$ are arbitrary integration constants. As a particular example we explicitly present the solutions corresponding to the case $n=2$, for which the quantum potential is
\begin{equation}
V(x)=f_0x^2+E\mp\sqrt{f_0},
\end{equation}
corresponding, from a physical point of view,  to  harmonic motion \cite{Merz}. The solution of the Schr\"{o}dinger - Riccati Eq.~(\ref{s1}) for this potential is
\begin{equation}\label{ref1}
u_{\pm}=\pm \sqrt{f_0}x+\frac{e^{\pm \sqrt{f_0}x^2}}{C_{\pm}-\sqrt{\pi }F_{\pm}\left(f_0^{1/4}x\right)/f_0^{1/4}},
\end{equation}
where $F_{+}\left(z\right)={\rm erf}(iz)/i$ and $F_{-}\left(z\right)={\rm erf}(z)$, with ${\rm erf}(z)=\left(2/\sqrt{\pi}\right)\int _0^z{e^{-t^2}dt}$ \cite{Abram}. The wave function for the harmonic double well potential is obtained as
\begin{equation}\label{ref2}
\psi _{\pm }\left( x\right) =\psi _{\pm 0}\exp \left[ \mp \sqrt{f_{0}}%
x^2/2\right]\left[ C_{\pm }+\sqrt{\pi }F_{\pm}\left(f_0^{1/4}x\right)/f_0^{1/4}\right].
\end{equation}

Eqs.~(\ref{ref1}) and (\ref{ref2})  have been used in the physical context of quantum mechanics since 1984 \cite{Mi,Ro}.

As a second example of the application of the integrability condition given
by Eq.~(\ref{b3}) to the Schr\"{o}dinger - Riccati system we consider the
case $f(x)=f_{0}x^{n}-E$, $n\neq 0$. The corresponding potential, satisfying the
integrability condition of the Riccati equation, is
\begin{equation}\label{pot2}
V(x)=f_{0}x^{n}\mp \frac{f_{0}n}{2}\frac{x^{n-1}}{\sqrt{f_{0}x^{n}-E}},n\neq 0.
\end{equation}

The general solution of the Riccati equation Eq.~(\ref{s1}) for this
potential is
\begin{equation}
u_{\pm }(x)=\pm \sqrt{f_{0}x^{n}-E}+\frac{H_{\pm }\left( x\right) }{C_{\pm
}-\int ^x H_{\pm }\left( x'\right) dx'},  \label{uu}
\end{equation}%
By substituting Eq.~(\ref{uu}) into Eq.~(\ref{phi}), the wave function can
be represented by
\begin{equation}
\psi _{\pm }(x)=\psi _{\pm 0}H_{\mp }\left( x\right) \left[ C_{\pm }-\int ^x
H_{\pm }\left( x'\right) dx'\right] ,
\end{equation}%
where we have introduced the function $H_{\pm }\left( x\right) $, given by
\begin{equation}
H_{\pm }\left( x\right) =\exp \left\{ \pm 2x\sqrt{f_{0}x^{n}-E}\left[
2+n\;_{2}F_{1}\left( 1,\frac{1}{2}+\frac{1}{n};1+\frac{1}{n};\frac{f_{0}x^{n}%
}{E}\right) \right] /\left( 2+n\right) \right\} .  \label{psi2}
\end{equation}

In Eq.~(\ref{pot2}) for all $n \neq 0$  the
 potentials are singular at the points $f_0x^n = E$. This singularity can be avoided, from a physical point of view, by taking into consideration only  bound physical states, that is, those for which the energy $E$ is negative, and by considering only half-line
(radial) cases.

\subsection{The Navier-Stokes equation}

The Navier-Stokes equation for a steady viscous flow,
\begin{equation}
\rho \left( \vec{v}\cdot \nabla \right) \vec{v}=-\nabla p+\rho \vec{f}+\mu
\Delta \vec{v},
\end{equation}%
where $\rho $ is the density of the fluid, $p$ is the fluid pressure, $\vec{v%
}$ is the velocity, $\mu $ is the dynamic viscosity, and $\vec{f}$ is the
external force acting on the fluid, is one of the most complex equations of
mathematical physics. Therefore reducing it to a simpler form, or establishing, by using some physically reasonable assumptions, a connection between the Navier-Stokes equation and some other equations of the mathematical physics, is of fundamental importance in obtaining some exact solutions of the Navier-Stokes equation. By introducing a strain field $\gamma $ into the velocity field $\vec{u}$ of the fluid so that $\vec{u}=\left[-\left(\gamma/2\right))x,-\left(\gamma /2\right))y,\left(\gamma \right)z+\left(-\partial \psi/\partial y, \partial \psi /\partial x, W\right)\right]$, where $\psi $, $\gamma $ and $W$ are functions of $x$, $y$ and $t$, but not of $z$, solutions of the Burgers type can be constructed \cite{Gibbon}. If the strain rate $\gamma $ is a function of time only, $\gamma =\gamma (t)$, then it is related to the pressure $p$ of the fluid by a Riccati equation of the form \cite{Gibbon}
\begin{equation}
\frac{d\gamma }{dt}+\gamma ^2+p_{zz}(t)=0,
\end{equation}
where $p_{zz}(t)$ is the second partial $z$-derivative of the pressure, which must be spatially uniform, a constraint
necessary for the existence of stretched vortex solutions of the 3D Navier $–$ Stokes  equations with uni-directional vorticity, which is stretched by a strain field that is decoupled from them. Therefore, if there is a function $f(t)$ so that the pressure can be represented as
\begin{equation}
p_{zz}(t)=\mp\frac{d}{dt}\sqrt{f(t)}-f(t),
\end{equation}
the strain rate $\gamma (t)$ for a viscous fluid flow can be obtained as
\begin{equation}
\gamma _{\pm }(t)=\pm \sqrt{f(t)}+\frac{d}{dt}\ln \left[ \Gamma _{\pm }+\int {e^{\pm
2\int ^t\sqrt{f(t')}dt'}dt'}\right] ,
\end{equation}
where $\Gamma _{\pm}$ are arbitrary constants of integration.

In \cite{Argentini} and \cite{Chris} it was shown that
along a streamline the two-dimensional Navier-Stokes equation can be written
as a Riccati equation of the form
\begin{equation}  \label{fluid}
\dot{u}_{1}-\alpha (s)u_{1}^{2}+\beta (s) =0,
\end{equation}%
where the dot denotes the derivative with respect to the parameter $s$, $%
\alpha =1/2\nu $, where $\nu =\mu /\rho $ is the kinematic viscosity, and $%
\beta =-(1/\nu )\left( \dot{q}/\rho -f_{1}\right) s+C/\nu $, with $C$ an
arbitrary constant. The parametrization of the trajectory is of the form $%
\Phi :s\rightarrow \left( \phi _{1}(s),\phi _{2}(s)\right) =\left(
x,y\right) $, and $\vec{u}=\vec{v}\circ \phi $ and $q=p\circ \phi $,
respectively. A solution of the Riccati equation in terms of the Airy
function was obtained in \cite{Argentini}. By using the integrability
condition Eq.~(\ref{b3}) more general solutions of the two-dimensional
Navier-Stokes equation can also be constructed. Hence, if there exists a function $f(s)$ so that the condition
\begin{equation}
\pm \frac{d}{ds}\sqrt{\frac{f(s)}{\alpha (s)}}=-\beta (s)+f(s),
\end{equation}
is satisfied for all $\alpha (s)$, $\beta (s)$ and $f(s)$, the general solution of Eq.~(\ref{fluid}) is given by
\begin{equation}
u_{\pm 1}(s)=\pm \sqrt{\frac{f(s)}{\alpha (s)}}+\frac{e^{\pm 2\int ^s \sqrt{f\left(s'\right)\alpha \left(s'\right)}%
ds'}}{U_{\pm }-\int ^s\alpha \left(s'\right)e^{\pm 2\int ^{s'} \sqrt{f\left(s''\right)\alpha \left(s''\right)}ds''}ds'},  \label{th2}
\end{equation}%
where $U_{\pm }$ are arbitrary integration constants.

\section{Conclusions}

\label{sect4}

In the present paper, by extending the work initiated in \cite{AMC}, we have
obtained two integrability conditions for the Riccati equation, one for the
"full" equation, and one for its "reduced" form, respectively. Both
integrability cases are based on the correspondence between the initial
Riccati equation and a more general equation containing a solution
generating function $f(x)$. If the coefficients of the Riccati equations and
the function $f(x)$ satisfy some differential integrability conditions, the
general solutions of the considered Riccati equations can be explicitly
obtained. However, we would like to mention that generally the linear term in the Riccati equation, containing the function $b(x)$, can be eliminated from the initial Riccati equation by means of the transformation
\begin{equation}
y(x)=e^{\int ^{x}b\left(x'\right)dx'}v(x),
\end{equation}
with $v(x)$ satisfying the "reduced" Riccati equation
\begin{equation}\label{trans}
\frac{dv}{dx}=a(x)e^{-\int ^{x}b\left(x'\right)dx'}+c(x)e^{\int ^{x}b\left(x'\right)dx'}v^2.
\end{equation}

The integrability conditions given by Eqs.~(\ref{th1}) and (\ref{th2}) can also be applied to Eq.~(\ref{trans}). However, sometimes in practical applications the analysis of the solutions of the Riccati equation can be done easier in the general form, especially in the situations in which the integral of the function $b$ cannot be obtained in an exact analytical form.

We have also presented some explicit physical applications of the
integrability conditions for the Schr\"odinger-Riccati and the
two-dimensional Navier-Stokes equations. Therefore, for the evolution
equation of natural processes that can be reduced to a Riccati form, the
presented integrability conditions open the possibility of finding explicit
exact solutions for physical models showing a complex dynamical behavior.

\acknowledgments We would like to thank to the anonymous referee, whose comments and suggestions helped us to significantly improve our manuscript.

\end{document}